# Modulating Curie Temperature and Magnetic Anisotropy in Nanoscale Layered Cr$_2$Te$_3$ Films: Implications for Room-Temperature Spintronics

*In Hak Lee[1,2], Byoung Ki Choi[1], Hyuk Jin Kim[1], Min Jay Kim[1,3], Hu Young Jeong[4], Jong Hoon Lee[4], Seung-Young Park[5], Younghun Jo[5], Chanki Lee[2], Jun Woo Choi[2], Seong Won Cho[6], Suyuon Lee[6], Younghak Kim[7], Beom Hyun Kim[8], Kyeong Jun Lee[9], Jin Eun Heo[9], Seo Hyoung Chang[9], Fengping Li[1], Bheema Lingam Chittari[1], Jeil Jung[1,3], Young Jun Chang[1,3,][*]*

[1]*Department of Physics, University of Seoul, Seoul 02504, Korea,*
[2]*Center for Spintronics, Korea Institute of Science and Technology, Seoul, 02792, Korea,*
[3]*Department of Smart Cities, University of Seoul, Seoul 02504, Korea,*
[4]*UNIST Central Research Facilities (UCRF), UNIST, Ulsan 44919, Korea,*
[5]*Center for Scientific Instrumentation, Korea Basic Science Institute, Daejeon 34133, Korea,*
[6]*Center for Electronic Materials, Korea Institute of Science and Technology, Seoul 02792, Korea,*
[7]*Pohang Accelerator Laboratory, POSTECH, Pohang, 37673, Korea,*
[8]*Korea Institute for Advanced Study, Seoul 02455, Korea,*
[9]*Department of Physics, Chung-Ang University, Seoul, 06974 Korea.*

**ABSTRACT**

Nanoscale layered ferromagnets have demonstrated fascinating two-dimensional magnetism down to atomic layers, providing a peculiar playground of spin orders for investigating fundamental physics and spintronic applications. However, strategy for growing films with designed magnetic properties is not well established yet. Herein, we present a versatile method to control the Curie temperature ($T_C$) and magnetic anisotropy during growth of ultrathin $Cr_2Te_3$ films. We demonstrate increase of the $T_C$ from 165 K to 310 K in sync with magnetic anisotropy switching from an out-of-plane orientation to an in-plane one, respectively, via controlling the Te source flux during film growth, leading to different c-lattice parameters while preserving the stoichiometries and thicknesses of the films. We attributed this modulation of magnetic anisotropy to the switching of the orbital magnetic moment, using X-ray magnetic circular dichroism analysis. We also inferred that different c-lattice constants might be responsible for the magnetic anisotropy change, supported by theoretical calculations. These findings emphasize the potential of ultrathin $Cr_2Te_3$ films as candidates for developing room-temperature spintronics applications and similar growth strategies could be applicable to fabricate other nanoscale layered magnetic compounds.

# INTRODUCTION

The discovery of graphene had prompted huge research efforts to explore the two-dimensional (2D) layered material family, such as graphene,[1] hexagonal boron nitride,[2] black phosphorus[3] and transition metal chalcogenides (TMCs).[4] In their atomically thin forms, 2D layered materials exhibit exceptional physical properties, such as superconductivity,[5] charge density wave,[6,7] valley dichroism,[8] tunable bandgap,[4] ferroelectricity,[9] and ferromagnetism.[10,11] In particular, 2D layered magnets have demonstrated fascinating 2D magnetism down to the atomic layers, providing a peculiar playground of spin orders for investigating fundamental physics and potential spintronic applications. Based on the weak layer interactions, it is possible to study the fundamental behaviors of 2D magnetism while avoiding effects of substrates or capping layers.[12] Furthermore, it is extremely important to correlate the magnetism of layered 2D heterostructures with various other physical properties to develop multifunctional device applications.[13]

Recently, exfoliation of 2D magnetic crystals has enabled us to identify long-range ferromagnetic (FM) ordering in ultrathin forms, such as insulating monolayer $CrI_3$,[10] insulating bilayer $Cr_2Ge_2Te_6$,[11] and metallic few layers $Fe_3GeTe_2$,[14] in which the magneto- crystalline anisotropy due to reduced crystal symmetry promotes long-range FM ordering.[15] Other than the exfoliated flakes of magnetic crystals, several new magnetic materials, such as $Fe_3GeTe_2$,[16] $MnSe_x$,[17] $VSe_2$,[6] $V_5Se_8$,[18] and $VTe_2$,[19] exhibit FM in their thin film forms, which enable their large-scale fabrication and applications. Owing to the extremely small volume of ultrathin films, different magnetic measurements, such as the magneto-optical Kerr effect, and anomalous Hall effects, X-ray magnetic dichroism (XMCD) measurements, are also employed to distinguish intrinsic ferromagnetism.[18,20] Till date, the Curie temperatures ($T_C$) observed in the flakes of 2D layered ferromagnets are significantly low; hence, large research efforts are focused on enhancing the $T_C$ above room temperature. Carrier-doping-dependent modification of the $T_C$ has been demonstrated via either electric-field or composition dependence in $Fe_3GeTe_2$ flakes.[21,22] Along with an enhanced $T_C$, controlling the magnetic anisotropy (MA) is important for realizing practical layered-material-based spintronics devices; MA is a critical material parameter that determines many spin transport characteristics. While perpendicular MA (PMA) materials are preferred for some spintronics applications owing to the improvement in storage density and reduced power consumption that PMA provides,[23] in-plane MA (IMA) materials are better suited for other applications, such as fast spin switching.[24] Nevertheless, there are only a limited number of studies reporting the control of MA in layered ferromagnets.[22,25] According to recent theoretical calculations, layered $Cr_{1+x}Te_2$ compounds may be one of candidate materials for realizing room temperature 2D FM states.[26] $Cr_{1+x}Te_2$ compounds, such as CrTe (x=1), $Cr_2Te_3$ (x=0.33), $Cr_3Te_4$ (x=0.5), $Cr_5Te_8$ (x=0.25), and $CrTe_2$ (x=0), can be viewed as having alternating layers comprising Cr and Te atoms where the number of Cr atoms in every second Cr layer is gradually reduced.[27] $Cr_{1+x}Te_2$ compounds shows various $T_C$ values depending on the Cr contents (i.e., $T_C$ = 240 K for $Cr_5Te_8$, 195 K for $Cr_2Te_3$, and 340 K for $Cr_3Te_4$). In particular, $Cr_2Te_3$ shows PMA in bulk[28,29] and has been most found phase as thin film forms in the literature, where the $T_C$ varies from 170 K to 200 K.[30–35] However, a controlled method for tailoring the $T_C$ and MA in $Cr_{1+x}Te_2$ thin films without restrictions in terms of the thickness and stoichiometry is still lacking.

Therefore, we herein present a film growth method for tuning the $T_C$ and MA in $Cr_2Te_3$ thin films while maintaining similar thicknesses and stoichiometries. We utilized a high vacuum co-evaporation system for growing $Cr_2Te_3$ thin films with a controlled relative Te flux, as illustrated in Scheme 1. Magnetic studies demonstrated these films to be ferromagnetic with strikingly high variations in their $T_C$; the films exhibited switching between PMA and IMA, depending on the Te flux. Using XMCD, we confirmed the modulation of the anisotropic magnetic properties in connection with the orbital magnetic moments. Careful structural and microscopic analyses showed that $T_C$ and MA were correlated to the c-lattice constants of the films. Our work demonstrates a versatile growth method for achieving the largest variation in the magnetic properties of layered chromium telluride thin films, which proves the possibility of fabricating layered material-based magnetic heterostructures operating near room temperature.

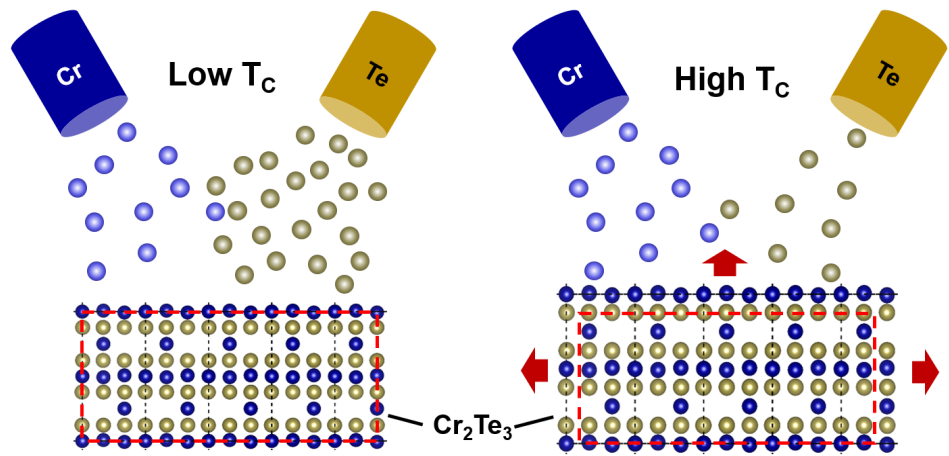

**Scheme 1.** Schematics of Te flux-controlled growth process for both low $T_C$ (high Te flux) and high $T_C$ (low Te flux) $Cr_2Te_3$ films.

**RESULTS AND DISCUSSION**

$Cr_2Te_3$ has a hexagonal NiAs structure with the space group $P\bar{3}1c$, as shown in Fig. 1(a). As observed in top view, the Cr and Te atoms form hexagonal patterns, similar to the $1T$ layered transition metal dichalcogenides. However, as observed in the side view, $Cr_2Te_3$ exhibits two orientations with distinct atomic arrangements resulting from to the intercalated Cr layers. In the $[11\bar{2}0]$ orientation, the Te atoms are vertically stacked with alternating layers of half-intercalated Cr atoms, whereas the Te atoms are stacked in zigzag manner with vertically aligned Cr lattices in the [10$\bar{1}$0] orientation. Such differences in stacking orders allowed us to identify in-plane crystalline orientations during microstructural analyses, as discussed later.

The $Cr_2Te_3$ thin films were grown on $Al_2O_3$ (0001) substrates in a high-vacuum co-evaporation system (see Methods section). $Cr_2Te_3$ films with modulated magnetic properties were grown by changing the Te source flux during film growth, as described in Scheme 1. The Te flux was varied so that the relative flux ratio of Cr:Te was 1:3.3 or 1:23. for low- and high- Te flux samples, respectively. (It is worth noting that the flux ratio is varied between 1:2 and 1:50 in existing studies, where the systematic influence on film properties is not adequately analyzed.[36,37]) Tuning the Te flux during film growth sharply changes the $T_C$. The temperature dependence of resistivity indicates metallic behavior, as shown in Fig. 1(b). Sudden slope changes are observed as ferromagnetic ordering emerges in the film below $T_C$, owing to reduced electron magnon scattering.[33] The $Cr_2Te_3$ films grown under the high Te flux exhibited a sudden decrease in resistivity below $T_C$ = 165 K (black and blue arrows, low $T_C$ (L$T_C$) samples), whereas the films grown under the low Te flux exhibited weaker slope changes near $T_C$ = 310 K (green and red arrows, high $T_C$ (H$T_C$) ones). Considering that ultrathin $Cr_2Te_3$ films grown at flux ratio of 1:7 showed varied $T_C$ at 165 K – 295 K [38], it is likely that the flux ratio with range of 1:3 – 1:7 would be near the critical value of Te flux for modulating magnetic properties in $Cr_2Te_3$ films. Moreover, we observed that the resistivity values remained similar for two different film thicknesses (i.e., 6.5 nm and 33 nm), thus we focused on the 6.5 nm thick films for investigating the differences in the magnetic properties of the H$T_C$ and L$T_C$ samples in this work. We note that small kink appears in the resistivity curve of H$T_C$ samples around 100 K, probably originated from a defect scattering, which have been also observed in the CVD-grown $Cr_2Te_3$ films[28].

Figure 1(c,d) shows the X-ray diffraction (XRD) patterns of the $HT_C$ and $LT_C$ films. Well-defined (002), (004), and (008) peaks of the films indicate that the trigonal structure of $Cr_2Te_3$ is well aligned along the [001] direction in both samples.[33] The weak (006) peak emerges for thicker films in the vicinity of strong $Al_2O_3$ substrate peaks. The thickness oscillations are evident near the (002) and (004) peaks, from which the thicknesses were confirmed to be 6.5 nm for both types of films. Figure 1(e) shows the c- and a-lattice parameters derived from the (004) and (202) peaks, respectively. The c-lattice values are estimated to be 12.20 – 12.21 Å (12.36 – 12.37 Å) for the $LT_C$ ($HT_C$) samples. Compared to the bulk lattice constant (12.07 Å),[39,40] the $HT_C$ samples have a more elongated c-lattice (2.45%) than the $LT_C$ ones (1.15%). The $HT_C$ samples also have a more elongated a-lattice (0.49%) than the LTC ones (-0.75%) compared to the bulk value (6.81 Å). Therefore, tuning Te flux stabilized the $Cr_2Te_3$ phases with different c-lattice and a-lattice values.

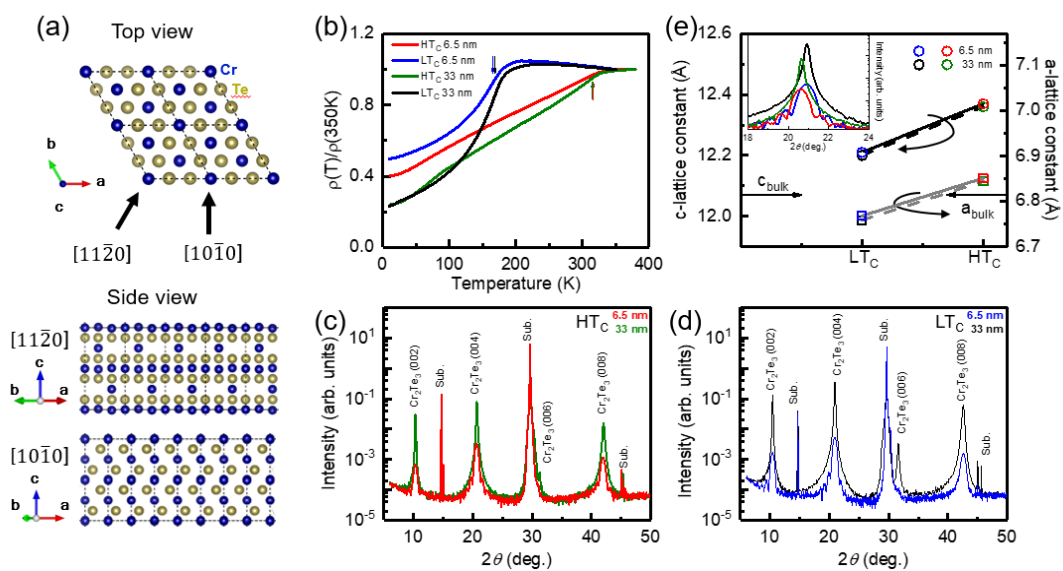

**Figure 1.** Structural and electrical property analyses of $Cr_2Te_3$ films grown on $Al_2O_3$ (0001). (a) Top and side views showing atomic structure of trigonal $Cr_2Te_3$ with the space group $P\bar{3}1c$. In the [11$\bar{2}$0] side view, alternating Cr and Te layers are stacked with Cr vacancies in every other Cr layer. In the [10$\bar{1}$0] side view, Te atoms are located in a zigzag manner, whereas Cr atoms are aligned vertically. (b) Temperature dependence of electrical resistivity normalized with the value at 350 K. Vertical arrows indicate the transition points where the slope of the resistivity curve changes owing to ferromagnetic phase transition for the samples with high (HT$_C$) and low $T_C$ (LT$_C$). (c,d) High-resolution XRD patterns for HT$_C$ and LT$_C$ samples. Each diffraction peak is indexed to the $Cr_2Te_3$ films and $Al_2O_3$ substrates. (e) Comparison of c- and a-lattice constants estimated from the $Cr_2Te_3$ (004) (inset) and (202) peaks, respectively. The horizontal arrows indicate the bulk c-lattice (12.07 Å) and a-lattice (6.81 Å) values.

Magnetization measurements revealed strikingly different magnetic properties with respect to the $T_C$ and MA in HT$_C$ and LT$_C$ samples. Figure 2(a,b) show temperature-dependent magnetization curves for the 6.5-nm thick films. Both films demonstrated clear ferromagnetic signals but with different orientations, whereas the onset temperatures are observed at ~300 K (HT$_C$) for the in-plane (IP) direction and at ~190 K (LT$_C$) for the out-of-plane (OOP) direction, consistent with the estimated $T_C$ values from the resistivity curves in Fig. 1(b). Also, along the hard axis directions, magnetization signal shows smooth temperature dependences with similar onset temperatures compared to the easy axis behaviors in each sample. For the HT$_C$ film, the reduced Te flux tuning indeed stabilized its room-temperature ferromagnetism, yielding a considerably enhanced $T_C$ value than its bulk counterpart (195 K).[28] Although such an enhanced $T_C$ has recently been observed in $Cr_2Te_3$ films, it was for a thick film (300 nm, $T_C$= 295 K)[38] or for films with randomly distributed flakes having 7.1 nm thickness ($T_C$= 280 K).[28]

The field-dependent magnetization curves at different temperatures demonstrate switching of the magnetic easy axis between the samples. Figure 2(c,e) show clear hysteresis only along the IP direction for the $HT_C$ film. Contrary to the large saturation field along the OOP direction with negligible hysteresis, clear temperature-dependent hysteresis curves are observed along the IP direction. The situation is clearly opposite for the $LT_C$ film, where clear hysteresis curves with strong temperature dependence are present only along the OOP direction (Fig. 2(d,f)). Here, we also notice small kinks at low field regions, which may be originated from either thickness inhomogeneity or canted spin orientations of Cr atoms.[34,41] Along the IP direction, the magnetic moment does not show saturation behavior up to 2 T (not saturated up to 6 T, not shown here). The angle-dependent magnetoresistance (MR) and Hall effect measurements were obtained for the 6.5-nm-thick samples at 50 K confirmed that the magnetic easy axis of the $HT_C$ ($LT_C$) film was aligned along the IP (OOP) direction (see Fig. S1 in the Supporting Information). Such drastic differences in the MA have been previously reported only for $Cr_{1+x}Te_2$ films with high stoichiometric variation ($x$ = 0.33 – 0.82).[37]

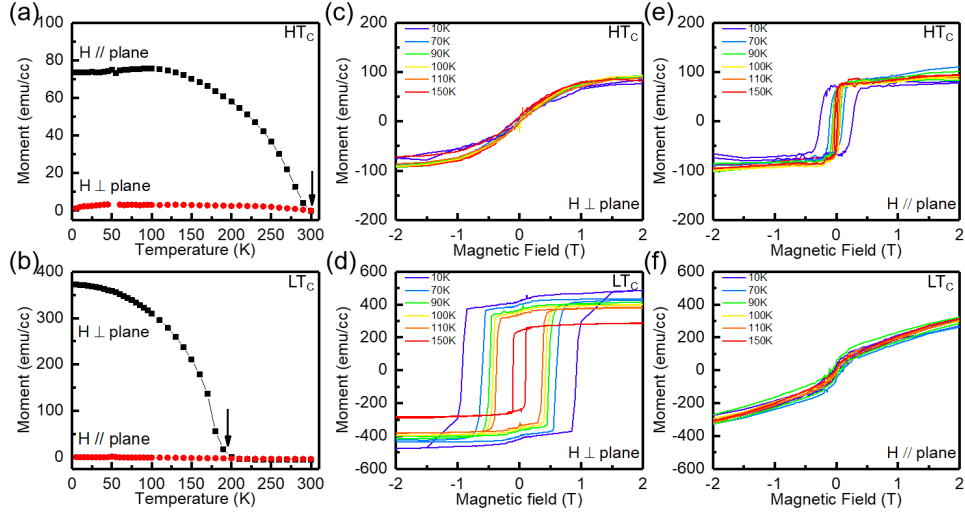

**Figure 2.** Magnetic properties of 6.5-nm-thick $Cr_2Te_3$ films. (a,b) Magnetization curves as a function of temperature under field cooling with a 100 Oe magnetic field ($H$) for $HT_C$ and $LT_C$ samples. Red and black data points were obtained with $H$ along OOP and IP directions, respectively. (c-f) Magnetization as a function of applied field at various temperatures (10 K – 150 K) with $H$ along OOP and IP directions, respectively.

To determine the intrinsic origins of the magnetism in the $Cr_2Te_3$ thin films, X-ray absorption spectroscopy (XAS) and x-ray magnetic circular dichroism (XMCD) were performed at the Cr $L$-edge for various temperatures and geometries. Figure 3 presents the spectra across the Cr $L_{2,3}$-edges and Te $M_{4,5}$-edges; however the integrated XAS intensity ratio of Te $M_{4,5}$/Cr $L_{2,3}$ was estimated to be 0.075, resulting in negligible contributions from the Te $M$-edge state.[42,43] Figure 3a compares the experimental spectra at 110 K and those obtained from the charge transfer multiplet calculations (see the details in the Methods section). The XAS spectra of the $HT_C$ and $LT_C$ films exhibit very similar shapes with the Cr $L_3$-edge peaks at 575.5 eV, consistent with the previously reported values for the bulk $Cr_2Te_3$ phase. By adopting parameters from previous reports, our calculated spectra also well matched well with the experimental XAS and XMCD spectra.[42,43] The spectral similarities between the HT$_C$ and LT$_C$ films indicate that the ferromagnetic signals originated from the intrinsic spin polarization of the Cr 3*d* electrons, and the chemical states of Cr in both films are considerably similar. Temperature-dependent XMCD measurements with two different magnetic field directions ensure the distinct MA states between the two samples. The XMCD signal is calculated from the difference in the XAS spectra taken with circularly polarized incident light under the field parallel ($\mu^+$) and field anti-parallel ($\mu^-$) conditions. As shown in the top parts of Fig. 3(b-e), clear differences in the XAS spectra obtained under $\mu^+$ and $\mu^-$ at 110 K indicate the existence of ferromagnetic signals regardless of the measurement geometries. As shown in Fig. 3(b,c), the HT$_C$ film exhibits robust XMCD signals up to 300 K along the OOP and IP directions, whereas the IP signal intensity further increased when cooled, consistent with the magnetization data. However, for the LT$_C$ film in Fig. 3(d,e), the XMCD signals emerge and gradually increase when the film is cooled below 200 K in both orientations. Although the temperature dependence of the OOP signals is consistent with the magnetization data, the IP magnetization also shows temperature independence. We noted that a remnant signal of the OOP component existed in the IP XMCD geometry, because the angle between the magnetic field and the sample plane was set to 30°, as shown in the inset of Fig. 3(e).

To deepen our understanding of the MA in the system, we evaluated the magnetic anisotropy energy (MAE), describing the magnetization tendency to align along certain orientations. The MAE is calculated from the integrals from the difference between the magnetization hysteresis loops measured at different orientations shown in Fig. 2(b-e).[37,44] The LT$_C$ sample shows a positive MAE (8.54 Merg/cc) indicating PMA, and the HT$_C$ one shows a negative MAE (-0.52 Merg/cc) indicating IMA. According to the Bruno model, the MAE is proportional to the difference between the m$_L$ values along the easy and hard axes.[45] Through analyzing the XAS and XMCD data using the sum rules[46], we estimated the relative ratio between the orbital and spin angular momentum ($m_L$/$m_S$), as -0.094 ± 0.009 (OOP) and -0.055 ± 0.005 (IP) for HT$_C$ and -0.043 ± 0.004 (OOP) and -0.107 ± 0.01 (IP) for LT$_C$, which are in the range reported previously.[37,43,47] As the moment ratio shows significant variations between the samples in the opposite direction, we estimated the anisotropy of the orbital moment as Δ(m$_L$/m$_S$)= (m$_L$/m$_S$)$_{OOP}$ − (m$_L$/m$_S$)$_{IP}$.[48,49] As shown in the inset of Fig. 3(a), the HT$_C$ (LT$_C$) has a negative (positive) value of Δ(m$_L$/m$_S$), exhibiting a trend similar to that observed for the change in the MAE, which is qualitatively consistent with the Bruno model. This suggests that the orbital moment plays an important role in steering the orientation of the spin moment, in turn modulating the MA via spin-orbit coupling.[48]. We also note that spin interactions in Cr-Cr or Cr-Te-Cr schemes are considered as another factor that affect the spin orientations in recent studies of Cr$_2$Te$_3$.[37,41]

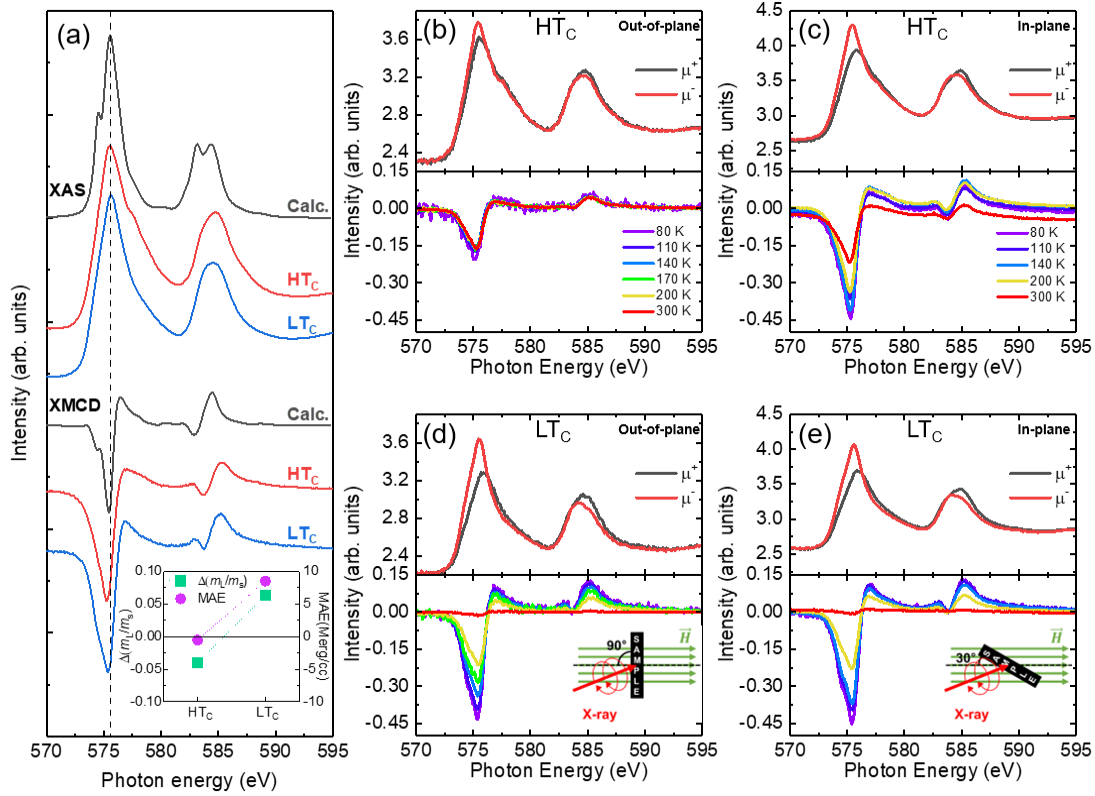

**Figure 3.** XAS and XMCD spectra of 6.5 nm-thick $Cr_2Te_3$ films. (a) Comparison of experimental (110 K) and calculated spectra for Cr $L_{2,3}$-edges. The simulated spectra were obtained from multiplet calculations of $Cr_2Te_3$ for XAS (average of field parallel ($\mu^+$) and antiparallel ($\mu^-$) to incident X-ray helicity) and XMCD ($\mu^+ - \mu^-$). Inset shows anisotropy of the orbital moment, $\Delta(m_L/m_S) = (m_L/m_S)_{OOP} - (m_L/m_S)_{IP}$, estimated from the XMCD at 110 K, *versus* MAE for Cr calculated from the magnetization hysteresis loops at 70 K shown in Fig. 2(c-f). (b-e) Experimental XAS and XMCD spectra for $HT_C$ and $LT_C$ samples with *H* along OOP and IP directions. While a single set of XAS spectra was chosen at 110 K for simplicity, XMCD spectra at various temperatures were determined (80 K – 300 K). Insets show the measurement geometries for the out-of-plane and in-plane settings.

Figure 4 shows the cross-sectional scanning transmission electron microscope (STEM) images of our films, revealing well-stacked layered structures. The bright field (BF)-

STEM images in Fig. 4(a,b) show atomically ordered patterns of both films on of the Al$_2$O$_3$ planes, where heavier elements are shaded darker. While the HT$_C$ film has continuous atomic layers with sharp a film substrate interface, the LT$_C$ has two types of atomic patterns, namely (i, iii) and (iii), with sharp lateral boundaries. The high-resolution high-angle annular dark field (HAADF)-STEM images demonstrate atomic arrangements with opposite contrast, as shown in Fig. 4(c-e), where the heavy Te atoms are clearly identified as bright balls and the Cr atoms are barely visible. When the substrate is rotated by 30° along the c-axis between the Al$_2$O$_3$[11$\bar{2}$0] and [10$\bar{1}$0] directions, the HT$_C$ film shows two dissimilar atomic patterns, corresponding to Cr$_2$Te$_3$ [10$\bar{1}$0] (Fig. 4(c)) and Cr$_2$Te$_3$ [11$\bar{2}$0] (Fig. 4(d)) crystal models, respectively. However, the Cr$_2$Te$_3$ [10$\bar{1}$0] and [11$\bar{2}$0] domains coexist in the LT$_C$ film, as shown in Fig. 4(e). Such coexisting 30°-twisted domains were previously observed in the Cr$_2$Te$_3$ film on Al$_2$O$_3$(0001), where interfacial amorphous layers (1–2-nm-thick) may allow such domains.[38] Similar twisted domain structures observed in hexagonal ZnO films grown on Al$_2$O$_3$(0001) have been suppressed through additional doping with vanadium during initial growth.[50] As LT$_C$ film shows interfacial amorphous regions, especially near the lateral boundaries between the domains (white arrows in Fig. 4(b,e)), we believe that a higher Te flux may change the adatom kinetics, which allows the initial formation of the twisted domains followed by different film growth modes.[51] To examine the microscopic stoichiometry of these films, energy-dispersive X-ray spectroscopy (EDS) was carried out on several different areas during the STEM measurements (see Fig. S2 in the Supporting Information). The HT$_C$ film showed a stoichiometric Cr$_{2+d}$Te$_3$ phase with $d = 0.01 - 0.02$, whereas the LT$_C$ film showed a rather larger variation with $d = -0.01 - 0.10$ among several regions. Although such stoichiometric variations are still small considering the effect of domain boundaries, one possibility is phase transition with different stoichiometries. It is known that both increase or decrease Cr content is expected to increase $T_C$ value (i.e., 340 K for $Cr_3Te_4$ or 240 K for $Cr_5Te_8$). However, the $HT_C$ films with higher $T_C$ show nearly same stoichiometry with $Cr_2Te_3$ while the $LT_C$ films with lower $T_C$ show larger stoichiometry variation from $Cr_2Te_3$.[37] Additional x-ray photoemission spectroscopy measurements confirm nearly same spectra for both samples, while the Cr and Te core level states are nearly overlapped in energy (see Fig. S4 in the Supporting Information). Therefore, these microscopic observations suggest that the excess Te flux does not significantly increase the Te content but modifies the film growth dynamics via changing the adatom mobilities and growth modes, which may kinetically stabilize different lattice parameter.[51] The resultant variation in the c-lattice constant may play a major role in changing the magnetic properties of these ultrathin films.

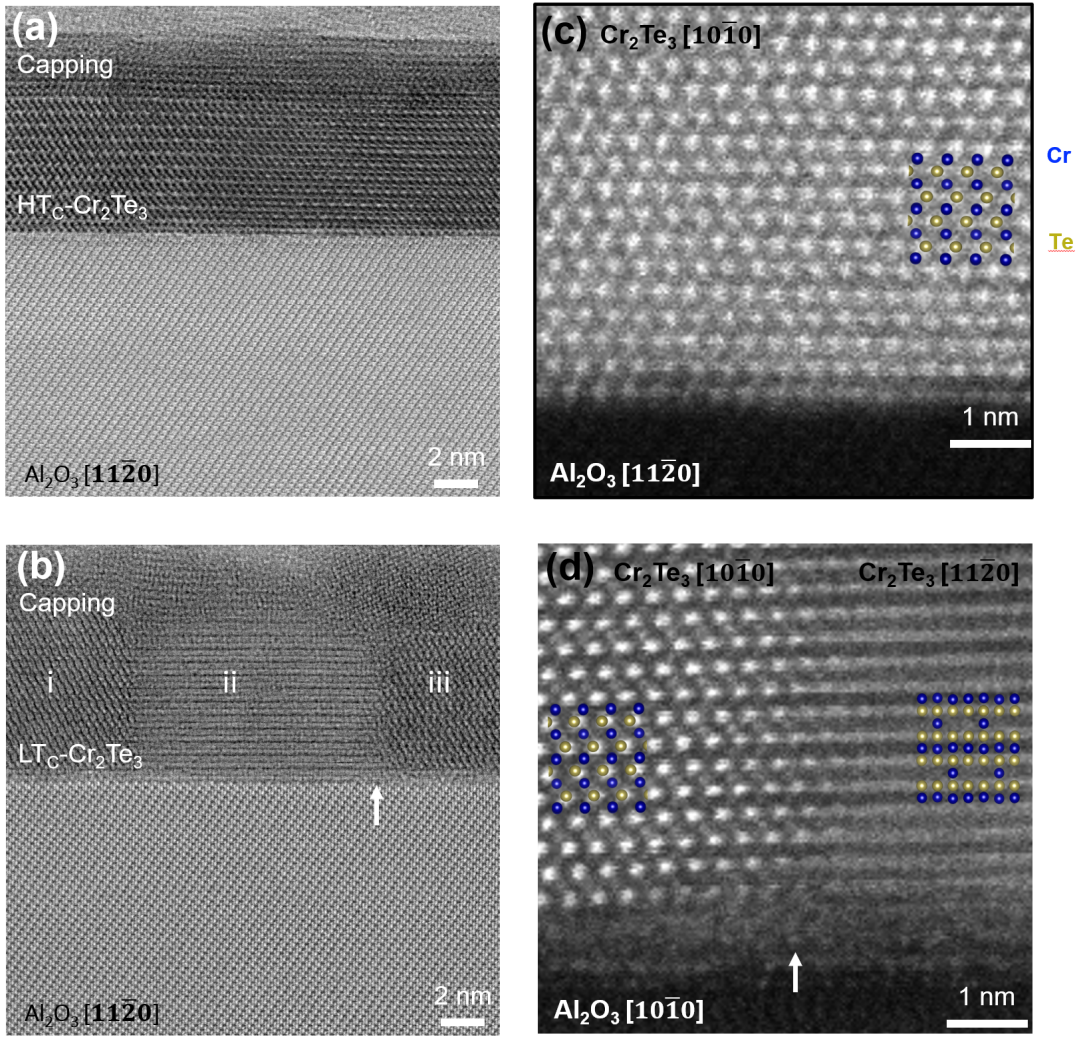

**Figure 4.** Cross-section STEM images and atomic arrangements of 6.5 nm-thick $Cr_2Te_3$ films. (a,b) BF-STEM images of (a) $HT_C$ and (b) $LT_C$ films showing smooth film thicknesses with sharp film-substrate interfaces. The $LT_C$ sample shows different atomic arrangements indicated as (i), (ii), and (iii), where (i) and (iii) have the same atomic registry. (c,d) High-resolution HAADF-STEM images of (c) $HT_C$ and (d) $LT_C$ films. White arrows in (b,d) indicate the lateral boundaries where interfacial amorphous layers are evident. Atomic models with different orientations are overlaid (Cr: blue spheres, Te: yellow spheres).

To deepen our understanding, we performed density-functional theory (DFT) calculations for the bulk $Cr_2Te_3$ compound. To simulate the experimental observation, we modulated the c-lattice parameters by varying them from 11.9 to 12.8 Å and fixing the lateral lattice parameters at a = b = 7.042 Å, which are obtained from the simulation result for fully relaxed structure of bulk $Cr_2Te_3$ (a=7.042 Å, c=12.568 Å) (see Table S1 in the Supporting Information). The theoretical MAE value ($MAE_0$) was calculated by subtracting the total energies between the systems with a given spin orientation and the energy of the spin parallel to the easy axis. As shown in Fig. 5(a), the $MAE_0$ has its minimum for the OOP ($\theta_0 = 90°$) configuration with spin ordering for relatively small c-lattice values. As the c-lattice parameter exceeds 12.4 Å, the minimum energy prefers the IP ($\theta_0 = 0°$) configuration. Figure 5(b) shows the gradual change in the $MAE_0$ as a function of the c-lattice parameter, demonstrating the switching of the MA from PMA to IMA. In the meantime, theoretically estimated $T_C$ values increased on increasing the c-lattice parameters, although the mean-field calculations of $T_C$ often overestimate the experimental observations owing to its simplified assumptions. These tendencies captured in the theoretical calculations well support the experimental observations, such as the changes in the magnetic easy axis and $T_C$. We also note that some discrepancies of $MAE_0$ and $T_C$ between experiment and theory may come from the sensitive influence of lattice parameters. Such consistent results further suggest the important role of the c-lattice constant in controlling the magnetic properties of these ultrathin $Cr_2Te_3$ layers, and possibly other related layered ferromagnets.

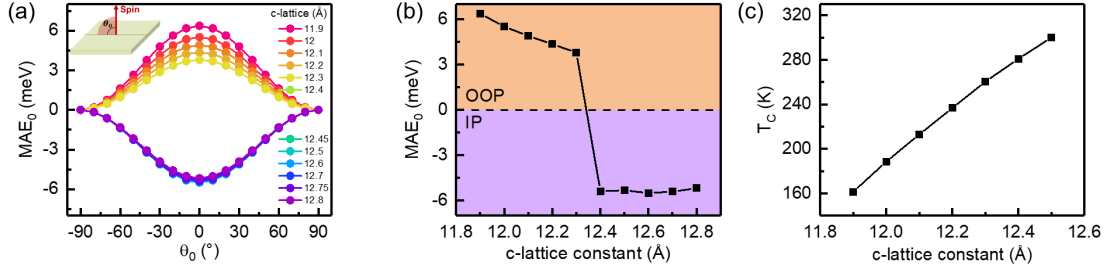

**Figure 5.** Theoretical calculation results of c-lattice dependence of $MAE_0$ and $T_C$. (a) $MAE_0$ as a function of $\theta_0$ for different c-lattice constants range from 11.9 Å to 12.8 Å. (b) $MAE_0$ as a function of c-lattice constant. Increasing the c-lattice constant flips the magnetic easy axis from the OOP ($\theta_0 = 90°$) to the IP ($\theta_0 = 0°$) direction. Orange (positive $MAE_0$) and purple regions (negative $MAE_0$) indicate the flipping of the easy axis from the c-axis to the ab-plane. (c) $T_C$ as a function of c-lattice constant. Increasing the c-lattice constant enhances estimated value of $T_C$.

Finally, the ultrathin layered ferromagnetic films with modulated magnetic properties obtained from easily tunable growth parameters allow us to obtain a variety of magnetic heterostructure. The $HT_C$ layer with IMA allows the development of giant tunneling magnetoresistance devices operating at near ambient conditions.[52] Further, the $LT_C$ layer with PMA is attractive for analyzing the interesting physics behind the topological skyrmion states and technological applications involving magnetic tunneling junctions, spin transfer torque, and tunneling magnetoresistance devices.[31,32,53,54] Our simple approach with a growth temperature below 400°C is widely adaptable to various types of heterostructure integration and device fabrication processes with similar chalcogenide magnetic compounds, such as topological insulators with large spin-orbit coupling and layered transition-metal dichalcogenides with large interface hybridization.[32,48] It is also extremely important to modulate the magnetic properties of Cr$_2$Te$_3$ films when grown on different substrates, which will pave the way to incorporate other complex materials into spintronics devices. [31,33,36,55]

**Conclusions**

In conclusion, we grew ultrathin Cr$_2$Te$_3$ thin films on Al$_2$O$_3$ substrates by tuning the Te flux ratio during high-vacuum co-evaporation. The $T_C$ and MA could be modulated for films with the same thickness and stoichiometries. The observed $T_C$ changed drastically from 165 K in the PMA film grown under the high Te flux to 310 K in the IMA film grown under the low Te flux for the same thickness. Magnetization and XMCD analyses results indicate that the modulated ferromagnetism is intrinsic and is related to the orbital magnetic moment. STEM analyses results reveal that the different Te fluxes influence the different growth dynamics in relation to different c-lattice values. DFT calculations confirm such contrasting behaviors under PMA and IMA depending on the c-lattice value, along with an increasing $T_C$. Therefore, the Te flux-controlled growth method will promote the development of new 2D ferromagnetic compounds and developing new layered magnetic device platforms operating at ambient conditions.

**Experimental Details**

**Film growth.** Cr$_2$Te$_3$ thin films were grown on Al$_2$O$_3$ substrates (0001) in a high-vacuum co-evaporation chamber with a base pressure lower than $5 \times 10^{-8}$ Torr. Prior to the growth of Cr$_2$Te$_3$, the substrate was heated to 400°C for 4 h, and then the Cr and Te sources were supplied for 1 h while keeping the substrate at 350°C under the shutter of the chamber for minimizing contamination. Cr (99.995%) and Te (99.999%) were evaporated using an e-beam evaporator and effusion cell, respectively, and co-deposited on the substrate at 350°C. During the deposition, the Te flux was set to 0.1

Å/s or 0.7 Å/s with respect to the Cr flux of 0.03 Å/s, so that the relative flux ratio of Cr:Te was 1:3.3 or 1:23 for the HT$_C$ and LT$_C$ samples, respectively. The thin film growth rate was approximately 0.8 nm/min, estimated from TEM analysis. The film deposition was followed by post-annealing at 400°C for 30 min. We deposited 5-nm-thick Pd or 2 – 10-nm-thick Al capping layers at room temperature to protect the film from surface oxidation.

**Structural and electrical characterizations.** High-resolution x-ray diffraction (XRD) measurements were performed using synchrotron radiation at the 3A MP-XRS (λ ~ 0.11145 nm, energy ~ 11.125 keV at Si(111)) beamline of Pohang Light Source-II (PLS-II, Pohang, Republic of Korea), and using an in-house XRD (Rigaku SmartLab) with 9 kW Cu K radiation (λ ~ 0.12398 nm). The electrical resistivity was measured as a function of temperature using a closed-cycle cryostat (Janis, CCS-22). Measurements were carried out in the van der Pauw geometry with square samples (5 mm × 5 mm) and indium Ohmic contacts in the sample corners. The four-terminal resistance was measured using a current source (224, Keithley Inc.) and a nanovoltmeter (2182, Keithley Inc).

**Magnetic measurements.** The magnetization measurements were carried out using a superconducting quantum interference device magnetometer (SQUID-VSM, Quantum Design Inc.) Two modes (zero field cooling and field cooling) were used for temperature-dependent magnetization measurements with a fixed magnetic field of 100 Oe. To investigate the MA of the Cr$_2$Te$_3$ thin films, angle-dependent magnetoresistance measurements were carried out in a commercial cryogen-free cryostat (Cmag Vari. 9, Cryomagentics Inc.) using the van der Pauw geometry with indium electrical contacts at four corners of the square samples. Electrical resistance was measured using a source-measure unit (2612A, Keithley Inc.) and a nanovoltmeter (2182, Keithley Inc).

**XAS/XMCD measurements and simulation.** To investigate the Cr L-edge state and the intrinsic magnetism in the $Cr_2Te_3$ thin films, temperature-dependent X-ray absorption spectroscopy (XAS) and X-ray magnetic circular dichroism (XMCD) were carried out at the 2A MS beamline at PLS-II. The total electron yield mode with an energy resolution of ~0.1 eV was used for both measurements at a base pressure of $5 \times 10^{-10}$ Torr in the analysis chamber by measuring the sample current ($I_1$) divided by the photon beam current ($I_0$) to remove the variations of beam intensity. To obtain the XMCD spectra, an external magnetic field of H = ±7000 Oe was applied with a fixed helicity of circularly polarized X-rays. The magnetic field directions were set to 0° and 60° with respect to the surface normal for the OOP and IP measurements, respectively. The angle of incidence of the photon beam was tilted at 22.5° from the magnetic field direction, as described in the inset of Fig. 3(d,e). To calculate the XAS and XMCD spectra, we employed the charge transfer multiplet calculation of $Cr^{3+}$ in the $O_h$ symmetry using the CTM4XAS package.[56] We assumed that the crystal field splitting between the $t_{2g}$ and $e_g$ orbitals, 10Dq, was 1 eV, Slater-Condon parameters were 70% of the atomic values, Coulomb repulsion between the valence $d$ orbitals $U_{dd}$ was 5 eV, and Coulomb repulsion between the valence $d$ and core $p$ orbitals $U_{pd}$ was 6 eV. To mimic the mixed valence feature of Cr in metallic $Cr_2Te_3$, the charge transfer energy $\Delta$ and hopping integral between $t_{2g}$ ($e_g$) and the ligand orbitals were set to 0 eV and 0.75 (1.5) eV, respectively. In this situation, the local ground multiplet state of Cr was considered as a coherent mixture of $Cr^{3+}$ $|\psi_{d^3}|{\sim}0.54$ and $Cr^{2+}$ $|\psi_{d^4L}|{\sim}0.46$.

**STEM measurements.** Cross-sectional specimens for scanning transmission electron microscopy (STEM) analysis were fabricated using a focused ion beam (FIB) technique (Helios Nano Lab 450, FEI) and additionally milled with a low-energy Ar-ion milling system (Fishione Model 1040 Nanomill). The STEM images were obtained using a double Cs-corrected FEI Titan³ G2 60-300 microscope with an accelerating voltage of 200 kV.

**Theoretical calculation method.** The total energy of ferromagnetic ($E_{FM}$) and antiferromagnetic ($E_{AFM}$) were obtained through plane wave density functional theory (DFT) calculations performed using the Vienna Ab initio simulation package(VASP).[57] The exchange-correlation functional was used semi-local PBE-GGA pseudopotentials with the vdW-D2 correction to account for vdW interactions.[58,59] The cutoff energy for the plane wave expansion of the DFT was 600 eV, and a 6 × 6 × 2 k-mesh was used in the Monkhorst−Pack method.[60] The conjugate-gradient (CG) algorithm was adopted to relax the atomic positions until the maximum atomic force acting on the relaxed ions was higher than 0.01 eVÅ$^{-1}$, and an energy break criterion of 10$^{-5}$ eV was applied for electronic optimization. The GGA+U approximation was considered for the correlation effects of the Cr 3d electrons, and U was set to 3.0 eV.[61,62] The lattice parameters of bulk Cr$_2$Te$_3$ were a = b = 7.042 Å and c = 11.9 – 12.8 Å. The lattice parameter of a = b = 7.042 Å is based on the simulation result for fully relaxed structure of bulk Cr$_2$Te$_3$ (a=7.042 Å, c=12.568 Å), which possess the lowest energy in comparison with other simulated results using different lattice parameters (see Table S1 in the Supporting Information). The mean field magnetic T$_C$ is expressed as follows[63,64]:

$$\frac{\gamma k_B T_C}{2} = E_{AFM} - E_{FM}$$

where $\gamma$ is the dimension of the system, i.e., 3, and $k_B$ is the Boltzmann constant.

The angular dependence of the MAE was analyzed using the plane wave DFT calculations implemented in Quantum Espresso (QE),[65] and we adopted the same convergence parameters as those used in the Curie temperature calculations. Firstly, we carried out static self-consistent calculations to obtain the wavefunction and charge density using the scalar-relativistic pseudopotential.[66] Secondly, the non-collinear calculation including the spin orbital coupling effect was performed using the previously obtained wavefunction and charge density with the fully-relativistic pseudopotential.

## ASSOCIATED CONTENT

Supporting Information

The Supporting Information is available free of charge on the ACS Publications website at DOI:

Further characterization on the MR, Hall effect, and EDS measurements.

The authors declare no competing financial interests.

## AUTHOR INFORMATION

**Corresponding Authors**

*E-mail: yjchang@uos.ac.kr.

**ORCID**

In Hak Lee: 0000-0001-6239-7101

Byoung Ki Choi: 0000-0003-3080-2410

Hyuk Jin Kim: 0000-0003-4620-9443

Min Jae Kim: 0000-0002-3086-098X

Hu Young Jeong: 0000-0002-5550-5298

Young Jun Chang: 0000-0001-5538-0643

**Author's Contributions**

Y.J.C. designed experiments, I.H.L., B.K.C., H.J.K., M.J.K., K.J.L., J.H., and S.H.C. performed sample preparations and sample characterizations, H.Y.J. and J.H.L. carried

out TEM analysis, S.-Y.P., Y.J., C.L., and J.C. examined magnetic measurements, S.W.C. and S.L. carried out magneto-transport measurements, I.H.L., B.K.C., H.J.K., Y.K. and B.H.K. performed XAS/XMCD measurements and its simulations, F.L., B.L.C., and J.J. performed band structure calculations, I.H.L. and Y.J.C. analyzed the results and wrote the manuscript.

**ACKNOWLEDGEMENTS**

This work is supported by the National Research Foundation of Korea (NRF) grants funded by the Korea government (NRF-2016M3D1A1900035, 2019K1A3A7A09033388, 2019K1A3A7A09033389, 2019K1A3A7A09033393, 2020R1C1C1012424, 2020R1A2C200373211). S.-Y.P., Y.J. acknowledge the supports of the KBSI grant (C030210) and the National Research Council of Science & Technology (NST) grant (No. CAP-16-01-KIST) by the Korean government (MSIP). C.L. and J.W.C. was supported by the KIST Institutional Program and National Research Council of Science and Technology (no. CAP-16-01-KIST). B.H.K. was supported by KIAS Individual Grants (CG068702).